\documentclass[a4paper]{aa}
\usepackage{natbib}
\usepackage{epsfig}

\begin{document} 

\newcommand{\chisq}{\mbox{$\chi^2$}}
\newcommand{\es}{erg s$^{-1}$}                          %energy cgs
\newcommand{\halpha}{H$\alpha$}                   %H I recombination lines 
\newcommand{\hbeta}{H$\beta$}
\newcommand{\kms}{km~s$^{-1}$}       %km/s
\newcommand{\cmthree}{cm$^{-3}$}
\newcommand{\msun}{M$_{\odot}$} 
\newcommand{\xmm}{XMM-\emph{Newton}} 
\newcommand{\chandra}{\emph{Chandra}} 
\newcommand{\nh}{\mbox{$N({\rm H})$}}
\newcommand{\rhoph}{$\rho$~Oph}
\newcommand{\eltn}{Elias 29}
\newcommand{\eltf}{Elias 24}
\newcommand{\hlt}{HL~Tau}

\title{A survey for Fe 6.4 keV emission in young stellar objects in
  \rhoph: the strong fluorescence from \eltn}

\author{F. Favata\inst{1} \and G. Micela\inst{2} \and B. Silva\inst{1}
  \and S.  Sciortino\inst{2} \and M. Tsujimoto\inst{3}}

\institute{Astrophysics Division -- Research and Science Support
  Department of ESA, ESTEC, 
  Postbus 299, NL-2200 AG Noordwijk, The Netherlands
\and
INAF -- Osservatorio Astronomico di Palermo, 
Piazza del Parlamento 1, I-90134 Palermo, Italy 
\and
Department of Astronomy and Astrophysics -- Pennsylvania State
University, 525 Davey Laboratory, University Park, PA 16802, USA
}

\offprints{F. Favata,\\ Fabio.Favata@rssd.esa.int}

\date{Accepted 17 Dec. 2004}

\titlerunning{Fe 6.4 keV fluorescence in \eltn} \authorrunning{F.~Favata
  et~al.}

\abstract{We report the results of a search for 6.4 keV Fe fluorescent
  emission in Young Stellar Objects (YSOs) with measured accretion
  luminosities in the \rhoph\ cloud, using the existing \chandra\ and
  \xmm\ observations of the region. A total of nine such YSOs have
  X-ray data with sufficiently high $S/N$ for the 6.4 keV line to be
  potentially detected if present. A positive detection of the Fe 6.4
  keV line is reported for one object, \eltn, in both the \xmm\ and
  the \chandra\ data. The 6.4 keV line is detected in \eltn\ both
  during quiescent and flaring emission, unlikely all previously
  reported detections of 6.4 keV Fe fluorescence in YSOs which were
  made during intense flaring. The observed equivalent width of the
  fluorescent line is large, at $W_\alpha \simeq 160$ eV, ruling out
  fluorescence from diffuse circumstellar material. It is also larger
  than expected for simple reflection from a solar-composition
  photosphere or circumstellar disk, but it is compatible with being
  due to fluorescence from a centrally illuminated circumstellar disk.
  The X-ray spectrum of \eltn\ is also peculiar in terms of its high
  (ionized) Fe abundance, as evident from the very intense Fe\,{\sc
    xxv} 6.7 keV line emission; we speculate on the possible mechanism
  leading to the observed high abundance.   } \maketitle

\section{Introduction}
\label{sec:intro}

The wealth of X-ray observations of star-forming regions carried out
in the 1990's by ROSAT and ASCA have established young stars as bright
X-ray sources throughout their evolution, from the Class I stage,
where an accreting envelope is still present around the young
star, throughout the Class III (or WTTS) stage, where very little
circumstellar envelope remains and the star's photosphere is hardly
distinguishable from that of a more mature similar object. Evidence
for X-ray emission from true protostars (Class 0 objects) has come
more recently, and it is still circumstantial.

While most YSOs are too far away for their X-ray emission to be
studied with the current generation of X-ray high resolution
spectrographs (with \xmm\ and \chandra), using CCD-resolution spectra
the X-ray emission from YSO's has been modeled as thermal emission
from a hot plasma in coronal equilibrium, with higher characteristic
temperatures than observed in older and less active stars. Due to
their proximity to Earth the X-ray spectra from YSOs in both the
Taurus and \rhoph\ regions have been studied in some detail (e.g. the
L1551 region in Taurus studied with \xmm, \citealp{fgm+2003}, the
\rhoph\ region studied with \chandra, \citealp{itk2002} and \xmm,
\citealp{ogm2004}), and in all cases the X-ray spectra were well fit
with a hot thermal plasma. Embedded YSOs (Class I) and Classical T Tau
stars (CTTS, or Class II) have higher characteristic X-ray
temperatures than Weak-Line T Tau stars (WTTS, or Class III): in
\rhoph\ \cite{ogm2004} find characteristic X-ray temperatures ranging
between 2.0 and 6.0 keV for Class I and Class II sources, while Class
III sources have quiescent X-ray temperatures ranging between 0.3 and
2.5 keV. Best-fit coronal abundances are (in line with many coronal
sources) rather low, with typical best fit values around
0.2--0.3\,$Z_\odot$, with some cases (e.g. XZ~Tau in L1551,
\citealp{fgm+2003}) showing very low abundances, $Z<0.1\,Z_\odot$,
where $Z_\odot$ is the solar photospheric abundance.

One notable deviation from a pure thermal X-ray spectrum has been
detected in the X-ray emission of the YSO YLW16A in \rhoph: during a
large flare \cite{ikt2001} detected, in addition to the Fe\,{xxv}
complex at 6.7 keV, a well visible 6.4 keV line which they attributed
to fluorescence from ``cold'' (i.e. neutral, or in low ionization
states) Fe. Such a line is produced when hard X-rays photo-ionize cold
material close to the X-ray source, and it is therefore an useful
diagnostic tool of the geometry of the X-ray emitting source and its
surroundings. The 6.4 keV fluorescent Fe line has been detected in a
number of astrophysical sources, from massive stars to SNR to the Sun
itself during flares, in which case the fluorescing material is the
solar photosphere. In YSOs, in addition to the photosphere, also the
circumstellar material could be responsible for the fluorescence,
either from the circumstellar disk and its related accretion
structures, or from more diffuse material.

\cite{ikt2001} do not discuss in detail the possible location of the
fluorescing material, although, from the lack of a delay between
appearance of the 6.7 keV and of the 6.4 keV lines they infer that the
fluorescing material must be associated with the star and located at
less than 20 AU from the X-ray source. The detection of Fe 6.4 keV
fluorescent emission from YLW16A has so far remained unique, in the
domain of cool stars, until the detection of Fe fluorescent emission
in a number of flaring sources in the Orion cloud by \cite{tfg+2004}
(see also \citealp{tfg+2004a}). In all these cases the 6.4 keV
fluorescent line was detected only during intense flares.

We present in this paper a survey for 6.4 keV fluorescent emission in
X-ray bright YSOs in the \rhoph\ cloud with measured accretion
luminosity. The original aim of our work was to search for
correlations between the characteristics of fluorescent emission and
the accretion luminosity. Our search has resulted in the first
detection of 6.4 keV fluorescent Fe emission in the quiescent emission
from a YSO, \eltn. 

The present paper is structured as follows: Sect.~\ref{sec:obs}
discusses the observations employed in the paper and their analysis,
the sample of stars analyzed is described in Sect.~\ref{sec:samp},
while the results are presented in Sect.~\ref{sec:res}. \eltn\ and its
strong fluorescent X-ray emission are discussed in detail in
Sect.~\ref{sec:eltn}, followed by the general discussion in
Sect.~\ref{sec:disc}. The conclusions are presented in
Sect.~\ref{sec:conc}.

\section{The sample}
\label{sec:samp}

Our starting sample consists of all the YSOs in the \rhoph\ cloud
which have measured accretion rates from Br$\gamma$ luminosity in the
work of \cite{mhc98}. Being in the IR, the Br$\gamma$ proxy of
accretion luminosity is little affected by interstellar matter and
allows to study also embedded sources. 

From the parent sample we selected the sources which were visible as
X-ray sources in at least one of the four X-ray observations listed in
Table~\ref{tab:sample}, and which had at least 500 net X-ray counts in
the spectrum, allowing a reasonably detailed spectral analysis to be
performed. The total number of YSOs satisfying this criterion is 9, of
which 3 are Class I objects and 6 are class II.  Given that some stars
are present in both \xmm\ and \chandra\ observations, a total of 12
spectra were analyzed.

The sample is listed in Table~\ref{tab:sample}, showing the source
name, its accretion and bolometric luminosity, and whether it was
observed by \chandra\ or by \xmm.

\begin{table*}[htbp]
  \caption{The sample of X-ray bright YSOs analyzed. The accretion
    luminosity reported is from \cite{mhc98}. }
  \label{tab:sample}
  \begin{tabular}{lrrrrr}
Source name & Class & $\log L/L_\odot$ & $\log L_{\rm acc}/L_\odot$ &
\xmm & \chandra \\\hline
\eltn\            & I  &    1.68$^1$ &    1.25 &  y & y  \\ %
WL  6            & I  &    0.38$^1$ & $-$0.37 &  -- & y  \\ % in % Chandra Koyama X-ray dark -- S23_4
GSS 26           & I  & $-$0.15$^1$ & $-$0.57 &  y &  gap  \\
YLW 3B (SR 24S)  & II & 1.33$^4$ &    0.34 & y & --  \\ % Done
\eltf\           & II & 0.23$^2$ &    0.34 &  y & y  \\ % src
DoAr25           & II & $-$0.08$^3$ & $-$1.34 & y & --  \\
V852 Oph (SR 22) & II & $-$0.08$^4$ & $-$1.34 & y & --  \\
VSSG 27          & II & $-$0.96$^3$ & $-$0.86 &  -- & y \\ 
VSSG 28 (GSS 39) & II & 0.04$^3$  & $-$0.62 &  -- & y \\\hline
  \end{tabular}
\newline$^1$ from \cite{mhc98}; $^2$ from \cite{djw2003}; $^3$ from
\cite{bak+2001}; $^4$ from \cite{cml+95}; $^4$~source unresolved from
YLW~3B (see discussion in the text). 
\end{table*}

\section{Observations and data analysis}
\label{sec:obs}

The \rhoph\ cloud has been observed by most X-ray imaging telescopes
flown to date; in particular it has been the subject of two \xmm\ and
two \chandra\ deep observations, with different pointing directions
(although with some overlap). While data from two \chandra\ 
observations and one \xmm\ observations have already been published,
we have, to be able to search for the 6.4 keV line in the individual
spectra (and to ensure homogeneity), re-analyzed all 4 observations in
a consistent way. The observations analyzed are listed in
Table~\ref{tab:xrayobs}. No analysis of observation 1 has been
published yet in the literature, while results from an analysis of
observation 2 have been recently published by \cite{ogm2004}. Results
from observation 3 have been published by \cite{ikt2001} and
\cite{int+2003}, who also have also published results from observation
4.

\begin{table*}[htbp]
  \caption{The four X-ray observations of the \rhoph\ star-forming cloud analyzed in the present work.}
  \label{tab:xrayobs}
  \begin{tabular}{lllrrrl}
N. & Mission &  Date & Exp. & RA & Dec & PI \\\hline
1 & \xmm\ &  11-09-2000 & 50 ks & 16:25:17.0 & $-$24:16:48 &
F. Favata \\ 
2 & \xmm\ &  19-02-2001 & 50 ks & 16:27:26.0 & $-$24:40:48
& M. Watson \\ 
3 & \chandra\  & 14-04-2000 & 100 ks & 16:27:18.1 &
$-$24:34:21 & K. Koyama \\ 
4 & \chandra\  & 15-05-2000 & 100 ks & 16:26:35.3 &
$-$24:23:12 & M. Gagn{\`e} \\ 
%  \begin{tabular}{llllrrrl}
%N. & Mission & Obs. ID & Date & Exp. & RA & Dec & PI \\\hline
%1 & \xmm\ & & 11-09-2000 & 50 ks & 16:25:17.0 & $-$24:16:48 &
%F. Favata \\ 
%2 & \xmm\ & 0111120201 & 19-02-2001 & 50 ks & 16:27:26.0 & $-$24:40:48
%& M. Watson \\ 
%3 & \chandra\ & 200060 & 14-04-2000 & 100 ks & 16:27:18.1 &
%$-$24:34:21 & K. Koyama \\ 
%4 & \chandra\ & 200062 & 15-05-2000 & 100 ks & 16:26:35.3 &
%$-$24:23:12 & M. Gagn{\`e} \\ 
  \end{tabular}
\end{table*}

%While the data from some of the observations have already been
%published in some form in the literature, we have chosen to analyze
%the data from the sources of interest in an uniform way here, to allow
%an unbiased comparison of the results.

Both \xmm\ observations have been processed using the same approach as
described in \cite{fgm+2003}, to which the reader is refereed for
details.  Briefly, the observations have been fully reprocessed using
the latest version of the SAS software (6.0), filtering out periods of
high background ('proton flares'), and photon lists have been
extracted for each source and for a relevant, nearby background
region; spectra and light curves have then been produced.

\chandra\ observations have been obtained from the archive, and no
further processing has been done on the cleaned photon lists; using
CIAO threads, photon lists have been extracted for each source and for
a relevant, nearby background region; spectra and light curves have
then been produced.

For each of the X-ray bright sources (listed in
Table~\ref{tab:sample}) the spectra have been analyzed using
\textsc{xspec}, and have been fit (in the energy band 1.0--8.0 keV)
using single-temperature \textsc{apec} models with varying global
metal abundance including an absorbing column density modeled with a
\textsc{wabs} model. In all cases the best-fit values are been fully
compatible with the values reported by \cite{ogm2004} for observation
2, and by \cite{ikt2001} and \cite{int+2003} for observations 3 and 4.
The presence of a 6.4 keV line was assessed by adding a gaussian line
to the spectrum, with initial energy $E = 6.4$ keV and  fixed
width (10 eV, the value expected from the width of the unresolved
fluorescence lines), and determining (with the use of the $F$ test)
whether this resulted in a significant improvement in the fit.

\section{Results}
\label{sec:res}

Table~\ref{tab:xraypar} shows the results of the spectral fitting
performed on all the X-ray bright sources in \rhoph. All sources could
be adequately fit with a single temperature absorbed spectrum, whose
parameters are reported in the Table. 

\begin{figure*}[!tbp]
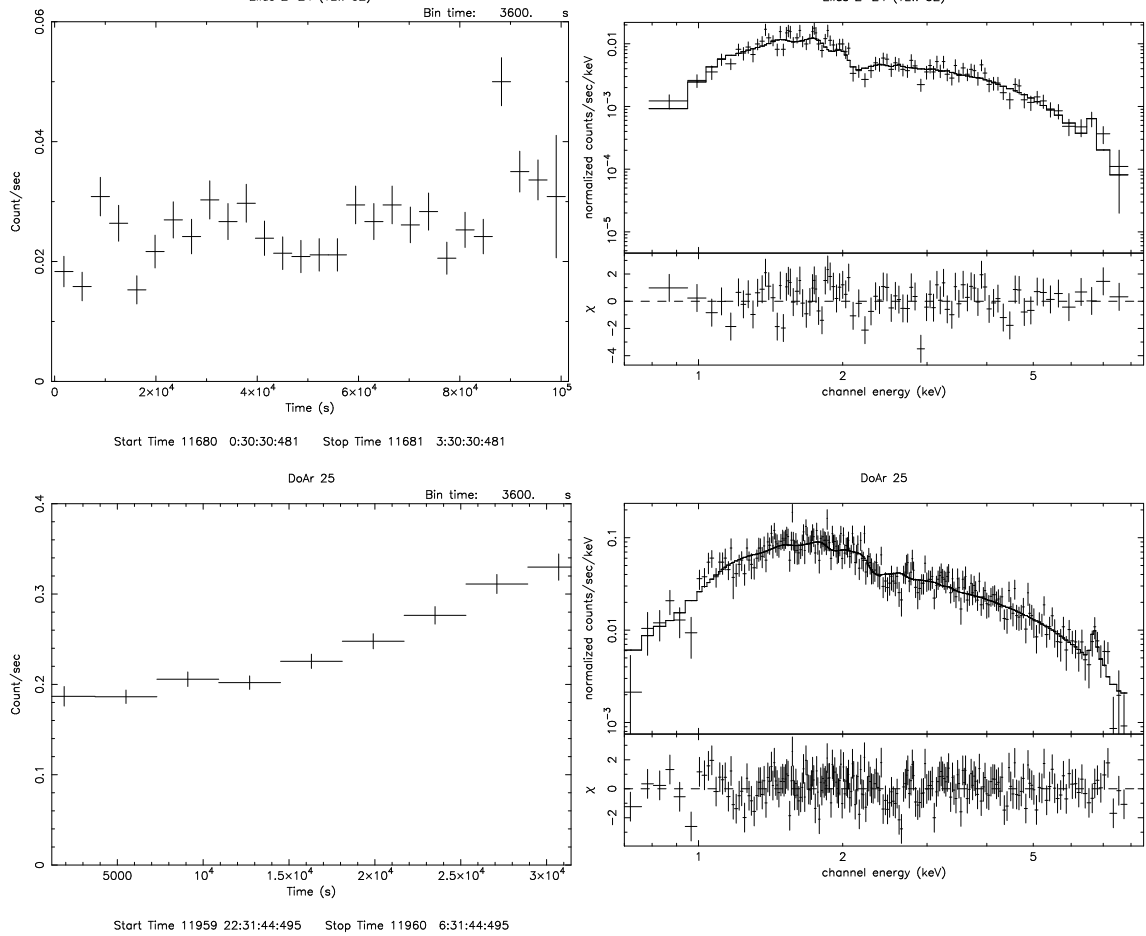

  \begin{center} 
    \leavevmode 
    \epsfig{file=2019f1.ps, height=7.5cm, angle=270}
    \epsfig{file=2019f2.ps, height=7.5cm, angle=270}\vspace{2ex}
    \epsfig{file=2019f3.ps, height=7.5cm, angle=270}
    \epsfig{file=2019f4.ps, height=7.5cm, angle=270}
  \caption{Top panels: Light curve and spectrum for the 
    \chandra\ observation of \eltf. Bottom panels: Light curve and
    spectrum for the \xmm\ observation of DoAr25.}
  \label{fig:el24}
  \end{center}
\end{figure*}

Fluorescent line emission at 6.4 keV was detected, using the procedure
described in Sect.~\ref{sec:obs}, only in \eltn; all other sources
show no significant evidence of excess emission above the purely
thermal spectrum. Two representative spectra and light curves (of the
stars \eltf\ and DoAr25) are shown in Fig~\ref{fig:el24}.

\begin{table*}[tbhp]
%  \centering
  \caption{Results of spectral fitting to the X-ray spectra of the
    sample stars. All fits where performed with a one temperature
    absorbed \textsc{apec} spectrum (fit results are given for a fit
    without a 6.4 keV line). Only \eltn\ showed evidence for
    the presence of 6.4 keV Fe fluorescence.}
\label{tab:xraypar}  
  \begin{tabular}{l| llll| llll}
& & \xmm & & & \chandra & \\
Source  & $T_{\rm X}$ & $L_{\rm X}$ & $Z$ & $\chi^2$ &
        $T_{\rm X}$ & $L_{\rm X}$ & $Z$ & $\chi^2$ \\
& (keV) & erg s$^{-1}$ & $Z_\odot$ & & (keV) & erg s$^{-1}$ &
$Z_\odot$ &  \\\hline
\eltn\  & $4.30 \pm 0.5$ &  $1.1\times 10^{30}$  & $1.06 \pm 0.18$ & 1.28 &
          $4.5 \pm 0.8$ &  $7.3\times 10^{29}$   & $0.60 \pm 0.14$ & 1.33 \\
WL 6    & -- & -- & -- & -- &
         $1.2 \pm 0.92$ & $5.2\times 10^{27}$   & 0.3 & 1.16 \\
GSS 26   & $4.82 \pm 1.1$ & $6.7\times 10^{29}$ &  $0.40 \pm 0.17$ & 1.19 
         & -- & -- & -- & -- \\%\hline
YLW 3B (SR 24S) & $2.69 \pm 0.33$ & $6.4\times 10^{29}$ & $0.16 \pm 0.15$ &
         1.12 & -- & -- & -- & -- \\
\eltf\  & $8.90 \pm 5.2$ & $8.6\times 10^{29}$ & 0.3 & 0.70 &
         $4.0 \pm 0.4$ & $1.2\times 10^{30}$ & $0.36 \pm 0.15$ & 1.24 \\
DoAr25   & $5.24 \pm 0.46$ & $1.4\times 10^{31}$ & $0.23 \pm 0.07$ & 0.97 
         & -- & -- & -- & --\\
V852 Oph & $0.74 \pm 0.15$ & $1.1\times 10^{29}$ & 0.3 & 0.74 
         & -- & -- & -- & --\\
VSSG 27 & -- & -- & -- & -- &
         $2.2 \pm 0.4$ & $5.5\times 10^{29}$ & 0.3 & 1.24\\
VSSG 28 & -- & -- & -- & -- &
         $1.8 \pm 1.3$ & $1.5\times 10^{29}$  & 0.3 & 1.05 \\

  \end{tabular}
\end{table*}

\subsection{\eltn}

\eltn\ is a well known luminous Class I YSO in the \rhoph\ cloud.
Using ISO observations \cite{bak+2001} determined its bolometric
luminosity at $L_{\rm bol} = 26\, L_\odot$, making it the most
luminous Class I source in \rhoph. \cite{mhc98} have used the emission
in the Br$\gamma$ line to determine the object's accretion luminosity
at $L_{\rm acc} = 15$--$18\, L_\odot$, making it the source with the
highest accretion luminosity in their sample.

Given the large veiling, very little information is available on its
characteristics; \cite{djw2003} obtained high-resolution IR spectra,
which are essentially featureless, indicating either a very large
amount of veiling, or very fast rotation (or both).

Using sub-millimiter observations \cite{bhc+2002} have separated the
contributions of the disk and of envelope surrounding the \eltn\ 
system, showing that the disk must be nearly face-on, with an
inclination $i > 60$ deg ($i = 90$ deg being face-on). Their
best-fitting disk model has an inner radius of 0.01 AU and an outer
radius of 500 AU, and a mass $M = 0.012\,M_\odot$. \cite{bhc+2002}
also argue that \eltn\ is a precursor of a Herbig AeBe star.

Using ISO and ground-based IR spectroscopic observations
\cite{cbt+2002} derive the presence of a significant component, in the
disk, of 'super-heated' gas, which, they argue, may be (partly) due to
the effects of high-energy (UV and X-ray) radiation from the central
source. 

\eltn\ has been observed in X-rays with ASCA, \xmm\ and \chandra, and
has been observed to flare repeatedly in X-rays (as common for YSOs),
e.g. by \cite{ikt2001}, who report flaring during the \chandra\ 
observation, and discuss the flares observed during the ASCA
observations. The temperature of the plasma during the \chandra\ 
observations was, according to \cite{ikt2001}, 4.3 keV in quiescence
and 7.5 keV during the flare, with respective X-ray luminosities of
$5.1 \times 10^{30}$ and $2.0 \times 10^{30}$ \es, with a fixed metal
abundance $Z = 0.3\,Z_\odot$. The \xmm\ observation of \eltn\ has been
previously reported by \cite{ogm2004}, who find coronal parameters
consistent with our own ($T = 4.3$ keV, $Z = 1.0 \, Z_\odot$).

%% Elias 29 observed by Barsony et al. (djw2003) but as it either
%% rotates too fast or it's too veiled they could not determine any
%% photospheric parameter.

\section{The X-ray characteristics of \eltn}
\label{sec:eltn}

\begin{figure*}[!tbp]
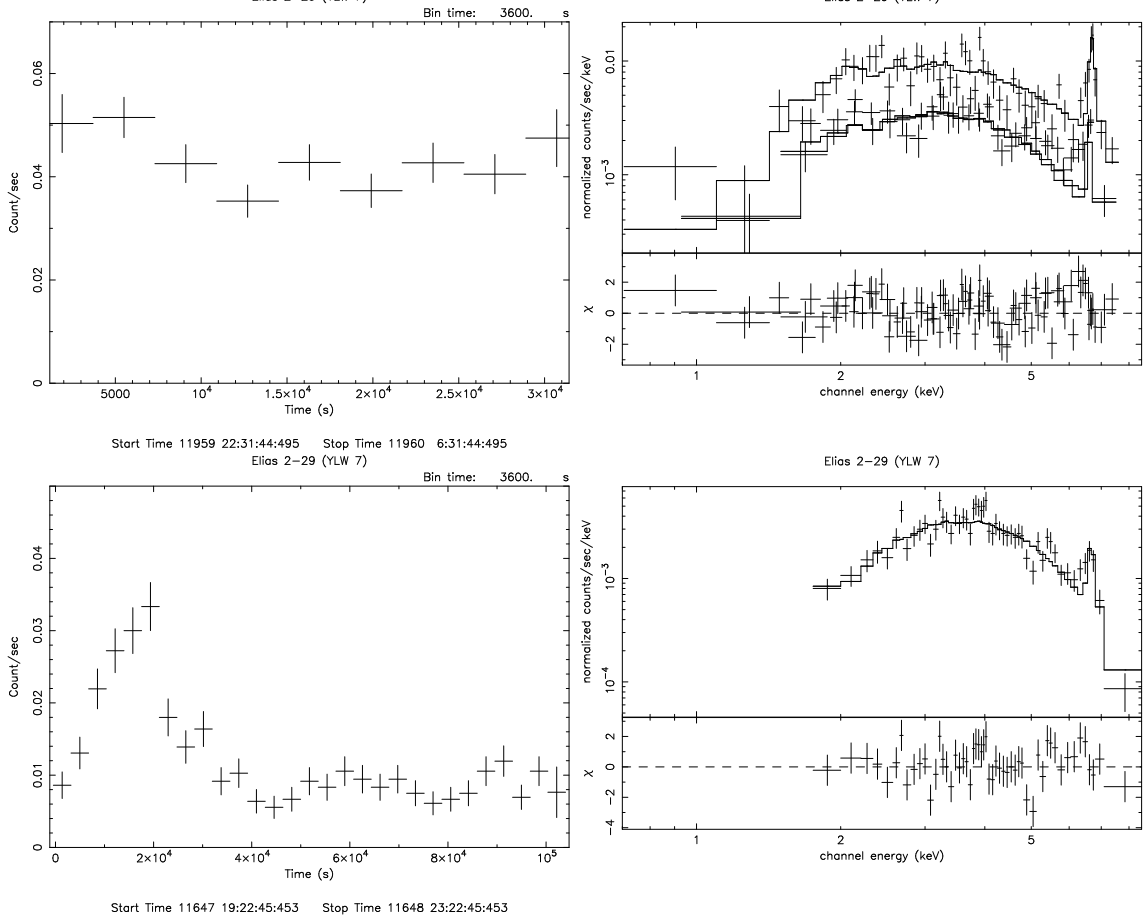

  \begin{center} 
    \leavevmode 
    \epsfig{file=2019f5.ps, height=7.5cm, angle=270}
    \epsfig{file=2019f6.ps, height=7.5cm, angle=270}
    \epsfig{file=2019f7.ps, height=7.5cm, angle=270}
    \epsfig{file=2019f8.ps, height=7.5cm, angle=270}
  \caption{X-ray light curves and spectra of \eltn. The \xmm\ light
    curve (pn only) and spectrum (for both the pn and MOS1 detectors)
    are shown in the upper panels, and the \chandra\ ones in the lower
    panels.}
  \label{fig:el29}
  \end{center}
\end{figure*}

The X-ray light curves (binned at 1 hr intervals) and spectra of
\eltn\ are shown in Fig.~\ref{fig:el29} for both the \xmm\ and
\chandra\ observations. While a significant flare was present in the
\chandra\ observation (as already reported by \citealp{ikt2001}),
during the \xmm\ observation the X-ray emission showed little
variability, with the source at an essentially constant level. Both
spectra are very hard, and with a very significant Fe 6.7 keV line
emission, which, for the \xmm\ EPIC pn spectrum, dominates even over
the lower-energy continuum.

A joint spectral fit to the \xmm\ pn, MOS1 and MOS2 data yields a best
fit temperature $T = 4.3 \pm 0.5$ keV, a coronal abundance $Z = 1.1
\pm 0.2~ Z_\odot$, and a colum density $N({\rm H}) = (4.8 \pm 0.3)
\times 10^{22}$ cm$^{-2}$. The X-ray luminosity, in the 1--7 keV band,
is $2.8 \times 10^{30}$ \es. A fit to the \chandra\ spectrum (which
has lower $S/N$ than the \xmm\ one) yelds a best fit temperature $T =
4.5 \pm 0.8$ keV, with a coronal abundance $Z = 0.60 \pm 0.15\,
Z_\odot$.  The column density is $N({\rm H}) = (7.4 \pm 0.5) \times
10^{22}$ cm$^{-2}$. The spectral parameters of the X-ray spectrum of
\eltn\ thus show little variation between the two observations.

While the high coronal temperature of \eltn\ is not exceptional, and
is rather typical of Class I YSOs, the coronal metal abundance
observed is very high (and the cause of such prominent 6.7 keV line
emission); typical values for YSOs are around 0.2--0.3 $Z_{\odot}$
(e.g. in \rhoph\ itself, \citealp{ikt2001}, \citealp{ogm2004} or in
Taurus, \citealp{fgm+2003}), and no other X-ray bright YSO shows such
high coronal metal abundance. While strong enhancements of the coronal
metallicity have been observed during flares (e.g. \citealp{fs99} in
Algol, or \citealp{tkm+98} in the YSO V773 Tau), the coronal
metallicity of \eltn\ is consistently high across the two
observations, and, if anything, higher during the (quiescent) \xmm\ 
observation than during the \chandra\ observation showing a flare.

\subsection{Fluorescent Fe K emission in \eltn}

\begin{figure*}[!tbp]
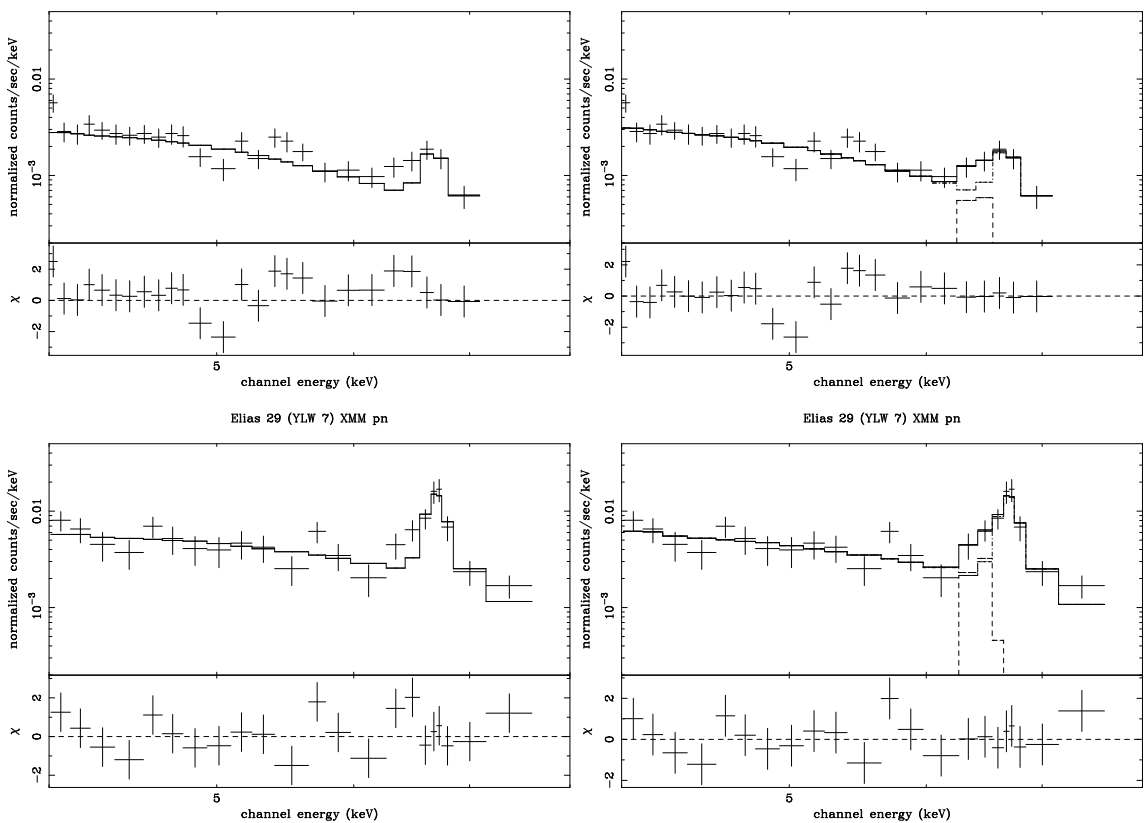

  \begin{center} 
    \leavevmode \epsfig{file=2019f9.ps, height=7.5cm,
      angle=270} \epsfig{file=2019f10.ps, height=7.5cm, angle=270}\vspace{2ex}
    \epsfig{file=2019f11.ps, height=7.5cm, angle=270}
    \epsfig{file=2019f12.ps, height=7.5cm, angle=270}
  \caption{The top spectra show the \chandra\ spectrum of \eltn\ in
    the 4.0--8.0 keV region, fit using an APEC thermal model (left)
    and with the addition of a narrow gaussian component (right). The
    bottom spectra show the same for the \xmm\ spectrum of \eltn.}
  \label{fig:el29fe64}
  \end{center}
\end{figure*}

A significant excess of emission redward of the prominent Fe K complex
at $\simeq 6.7$ keV, is visible in the \chandra\ spectrum of \eltn\ 
and (while less evident) is also well visible in the \xmm\ spectrum.
This excess emission occurs at the expected position of the 6.4 keV Fe
fluorescent line; to determine whether indeed such emission is present
in the X-ray spectrum of \eltn\ we have fit again the spectra with an
additional gaussian line component, which was constrained to be narrow
(10 eV), while its wavelength and normalization were left free to
vary. The other parameters (temperature, abundance and normalization
of the thermal spectrum) were also left free to vary. The resulting
fits are shown in Fig.~\ref{fig:el29fe64}, with the relevant
parameters listed in Table~\ref{tab:fe64}. For the \xmm\ data, for
clarity, only the pn spectrum is shown in Fig.~\ref{fig:el29fe64},
although the fit has been performed on the joint pn+MOS spectra.

In both cases, as indicated in Table~\ref{tab:fe64}, the best-fit line
energy agrees closely with the energy of the Fe fluorescent feature,
with $E = 6.44 \pm 0.05$ keV for the \chandra\ spectrum and $E = 6.43
\pm 0.04$ for the joint \xmm\ MOS+pn spectrum. The equivalent width of
the line ($W_\alpha$) is also similar across the two observations, at
150 eV and 120 eV for the \chandra\ and \xmm\ spectra respectively.
The $F$ test applied to the \chisq\ resulting from the fits with and
without the additional gaussian line shows that the feature is
significant, with a null hypothesis probability of the line being the
result of random fluctuation of 10\% and 8\% for the \chandra\ and
\xmm\ spectra respectively. In both cases, leaving the line width as a
free parameter does not improve the fit significantly, and the fit
converges onto essentially the same parameters, indicating that the
feature present in the spectrum is compatible with being due to a
narrow line.

To better determine the significance of the line, we also performed a
joint fit to the \xmm\ pn+MOS and \chandra\ ACIS spectra. The fit was
performed (in the interval 4.0--8.0 keV) with the addition of a
normalization constant (left free as a fit parameter) to allow for the
difference in X-ray luminosity of the source between the \xmm\ and
\chandra\ observations. The $\chi^2$ of the joint fit without the
addition of the 6.4 keV line is 1.61 over 62 degrees of freedom
(resulting in a low null hypothesis probability of 0.16\%); the
addition of the 6.4 keV line results in a much improved $\chi^2$ of
1.21 over 60 degrees of freedom, corresponding to a much higher null
hypothesis probability of 12\%. This results in a very high
significance of the line, better than 1\% (again using the $F$ test)
and results in a 6.4 keV fluorescence line with an equivalent width of
160 eV, which we will adopt in the following as the average value of
the fluorescence from \eltn. We have also separately analyzed the
\chandra\ flaring and non-flaring segments, but did not detect any
difference regarding the presence of the fluorescent line emission.
Finally, we verified that also the \textsc{mekal} and \textsc{raymond}
model produce similar results to the \textsc{apec} model.

\begin{table}[htbp]
  \caption{Parameters of the 6.4 keV fluorescent line detected in the
    \chandra\ and \xmm\ spectra of \eltn. The line energy $E$ is in
    keV, the line intensity in $10^{-6}$ photons cm$^{-2}$ s$^{-1}$ and the
    equivalent width $W_\alpha$ in eV. $P$ is the probability that the
    detected line is due to a random fluctuation in the data.} 
  \label{tab:fe64}
  \begin{tabular}{rrrrr}
Observation & $E$ & $I$ &  $W_\alpha$ & P \\\hline
ACIS-I & $6.45 \pm 0.05$ & $(1.6 \pm 1.3)$ & 150 & 10\% \\
pn+MOS & $6.43 \pm 0.04$ & $(1.3 \pm 1.0)$ & 120 & 8\%\\
ACIS-I+pn+MOS & $6.43 \pm 0.03$ & $(1.8 \pm 0.8)$ & 160 & $<1$\%\\ 
  \end{tabular}
\end{table}

\section{Discussion}
\label{sec:disc}

Fluorescent emission from the Fe 6.4 keV line has thus far only been
observed, for coronal sources, in intense flares from YSOs, with the
initial detection in a flare in YLW16A by \cite{ikt2001}, followed by
the recent sample of YSOs in Orion (all in flaring state) by
\cite{tfg+2004} and \cite{tfg+2004a}. \eltn\ is the first (and thus
far only) coronal source for which fluorescent emission is clearly
present in the quiescent state, i.e.\ in the absence of any significant
flare. In fact, the presence of a moderate flare in the \chandra\ 
observation does not seem to influence the presence of the fluorescent
emission, whose characteristics appear rather stable in time.

The intensity of the fluorescent emission is linked to both the amount
of available photo-ionizing photons (for the Fe 6.4 keV line all the
photons harder than the relevant photo-ionization edge, $\chi = 7.11$
keV) and to the amount and geometry of the photo-ionized material. 

In a simple formalism for the optically thin case, as discussed by
\cite{tfg+2004}, the equivalent width of the 6.4 keV fluorescence line
will be, under the assumption of a thermal spectrum with a temperature
of a few keV illuminating the cold material,
\begin{equation}
  \label{eq:ew}
  W_\alpha {\rm [eV]} = 2.5 \Bigl({\Omega\over{4\pi}}\Bigr)
\Bigl({N^\prime_{\rm H}\over{10^{22}~{\rm cm}^{-2}}}\Bigr) 
\end{equation}
where $\Omega$ is the solid angle of fluorescing material illuminated
by the thermal X-ray spectrum, and $N^\prime_{\rm H}$ is the
equivalent column density of the fluorescing material. In the case of
\eltn, with $W_\alpha \simeq 160$ eV, 
\begin{equation}
  \label{eq:ew2}
N^\prime_{\rm H} \simeq 6 \times 10^{23} \times {4\pi \over \Omega}~~
{\rm cm}^{-2} 
\end{equation}

Given that the circumstellar column density toward the X-ray source
determined spectroscopically is $N_{\rm H} = 4 \times 10^{22}$
cm$^{-2}$, this rules out (similarly to the Orion flaring sources
discussed by \citealp{tfg+2004}) fluorescence from diffuse circumstellar
material. Also, the typical column density at which the ionizing
photons are absorbed is $N_{\rm H} \simeq 10^{24}$ cm$^{-2}$
(\citealp{bai79}), so that the large equivalent width observed points
to the presence of optically thick conditions.

Under optically thick conditions, the computation of the equivalent
width of the fluorescence line requires a detailed radiative transfer
treatment and the assumption of a well defined geometry for the
absorbing material and for its illumination. \cite{gf91} have
performed this computation for the case of an accretion disk
illuminated either from above or from the central accreting source. In
their treatment they assume illumination from an X-ray source with a
power law spectrum, with photon indices $\Gamma$ varying between 1.3
and 2.3. While the \eltn\ spectrum is well fit with a hot thermal
spectrum, its continuum spectrum (ignoring the strong Fe 6.7 keV line,
which however, having an energy below $\chi = 7.11$ keV, does not
contribute to excite the fluorescence line) is equally well fit with
an absorbed power law with $\Gamma = 2.6$, somewhat softer than the
softer case studied by \cite{gf91}, but still sufficiently close to
allow a qualitative comparison. For the case of a disk illuminated
from above, the \cite{gf91} computation shows that the equivalent
width of the emitted fluorescent line varies roughly linearly with the
photon index, and $W_\alpha \leq 100$ eV for $\Gamma > 2$. The emitted
equivalent widths are somewhat larger (under the assumption of
favorable viewing geometry) for the case of a centrally illuminated
accretion disk, with $W_\alpha \leq 150$ eV for $\Gamma > 2$,
compatible with the equivalent width measured in the \eltn\ spectrum.

The observed fluorescence line can thus be naturally explained as
fluorescent emission from a centrally illuminated disk, seen face-on,
a natural geometry if X-ray emission indeed is concentrated near the
star. This implies that the disk is 'bathed' in high-energy X-rays
emitted by the star, thus supporting the hypothesis of \cite{cbt+2002}
that the 'hot' component they observe in IR in the disk is indeed
heated by the stellar high-energy radiation. \cite{cct+2002} have
recently discovered calcite toward a protostar similar to \eltn, and
\cite{cct+2005} report that calcite is also present toward \eltn\ 
itself. Given that calcite, under terrestrial conditions, requires the
presence of liquid water to form \cite{cct+2002} suggest that the
calcite they observe may thus form on the surface of grains in the
accretion disk, where water ice heated by the X-ray radiation may
acquire an enhanced mobility. Again, the detection of fluorescent 6.4
keV radiation from \eltn\ shows that its accretion disk is indeed
immersed in high-energy X-rays, and thus supports the above hypothesis
for the formation of calcite.

The X-ray spectrum of \eltn\ is also peculiar in terms of its high
coronal abundance, as visually shown by the strongly dominating Fe
line in the spectrum. While the relationship between coronal and
photospheric abundances is still far from clear (also largely due to
the lack of photospheric abundance measurements for active stars), in
the coronal abundance has been observed to vary significantly during
large flares, in a variety of stars (e.g. Algol, \citealp{fs99}, EV
Lac, \citealp{fmr+2001}, the PMS V773 Tau, \citealp{tkm+98}). In all
cases, the abundance increases from the typical $Z \simeq
0.2$--$0.3\,Z_{\odot}$ 'quiescent' value to $Z \simeq Z_{\odot}$. On
the other hand, the X-ray spectrum of \eltn\ shows a high Fe abundance
in its quiescent emission, with little if any influence from the small
flare detected in the \chandra\ observation.

Other stars in our sample have comparable X-ray characteristic coronal
temperatures, so that, in principle, a comparable number of
photo-ionizing photons would be available to generate fluorescent
emission. Yet, none of the other stars show any evidence for
fluorescent emission, and they all show the usual low coronal
abundance. While the information available on the characteristics of
\eltn\ is limited, given its being highly embedded and its high degree
of veiling, it is the most luminous source in the sample, both in
terms of its intrinsic luminosity and of its accretion luminosity: at
$L_{\rm acc} = 15$--$18\, L_\odot$ its accretion flux is more than one
order of magnitude more luminous than any of the other stars discussed
here.

While no correlation can be claimed on the basis of just one object
showing a number of interesting characteristics (an unique intense Fe
6.4 keV fluorescence outside of flares and a very high accretion flux,
together with a high metal abundance in the X-ray emitting plasma) it
is tempting to speculate about the possible link among them, in terms
of an emission driven X-ray mechanism (rather than a purely coronal
one). The free fall speed for a normal (non-degenerate) star is too
small for the infalling material to be heated at the $\simeq 50$ MK
temperature observed in the X-ray spectrum of \eltn, so that the
simple shock heating at the accretion site which explains rather well
the UV excess and the emission lines observed in the optical spectrum
cannot be invoked in the present context. In the magnetospheric
accretion scenarios normally invoked to explain the phenomenology of
CTTS the plasma is channeled from the accretion disk onto the stellar
surface in magnetic flux tubes, with one foot anchored on the stellar
photosphere and the other onto the edge of the accretion disk. The
currently most accredited mechanism for the heating of coronae is
shearing of the footpoints of the loops induced by convective motions
at the photosphere, where the loops are anchored. This mechanism would
still operate on the stellar side of the magnetic flux tube funneling
the accretion; on the accretion disk side, the Keplerian rotation of
the disk would provide a natural shearing force on the different parts
of the loop's footpoint at different distance from the star. Thus, the
accreting plasma in the magnetic flux tube funneling the accretion
would be subject to the same heating mechanisms as the plasma in
normal coronal loops, and would efficiently emit X-rays. The key
difference would be the continuous flow of fresh material in the flux
tube, similarly to what happens during the rise phase of a flare due
to chromospheric evaporation. This would naturally justify the high
metal abundance for the X-ray emitting plasma in \eltn, similarly to
the short-term increase in coronal metallicity observed during stellar
flares, in which the coronal abundance (likely due to the rapid influx
of chromospheric and photospheric material into the magnetic loop)
briefly increases to photospheric values. In the case of flares, once
the flow is stopped, the fractionation mechanism rapidly brings the
coronal abundance back to its pre-flare value; in the mechanism for
the X-ray emission from \eltn\ speculated here, the sustained flow
from the accretion disk into the magnetic flux tube would maintain the
high abundance in the X-ray emitting plasma.

\section{Conclusions}
\label{sec:conc}

Our survey for Fe 6.4 keV fluorescent emission in YSOs with measured
accretion luminosity in \rhoph\ has yielded a positive detection in
the Class I YSO \eltn, for which the 6.4 keV line is detected in both
the \xmm\ and the \chandra\ spectra. While in previous detections of
Fe 6.4 keV emission in YSOs the sources are always in a flaring state,
we detect the 6.4 keV emission in \eltn\ also while its thermal X-ray
emission shows no temporal variability. The thermal X-ray spectrum of
\eltn\ has an average characteristics temperature of $\simeq 4.6$ keV,
high but not exceptional among YSOs. Its Fe abundance is on the other
hand very high, with $Z = 1.1$ in the \xmm\ observation.

The equivalent width of the 6.4 keV line observed in \eltn\ is large,
with an average $W_\alpha = 160$ eV. The strength of the line rules
out reflection or transmission from diffuse circumstellar material,
and required an optically thick reflector. However, reflection from a
photosphere or circumstellar disk of solar photospheric composition
illuminated from above cannot produce, given the X-ray spectrum of
\eltn, a line of this strength. The observed equivalent width can
however be explained in a scenario in which the circumstellar disk is
illuminated by a centrally placed X-ray source, if the disk is
observed face-on. This scenario naturally explains the observed
fluorescent emission in \eltn\ (in which IR observations also show the
disk to have a large inclination, $i > 60$ deg, compatible with its
being face-on) and may simply explain the paucity of YSOs in which 6.4
keV has thus far been observed as a purely geometric effect, its
detection requiring observation of a (low-probability) face-on system.

The high abundance of the X-ray emitting plasma of \eltn\ could also
be naturally explained if the plasma is confined, rather than in
'classic' coronal loops, in magnetic flux tubes with one foot on the
star and the other on the accretion disk (as expected in the
magnetospheric model of accretion). In this case, material would
continuously flow within the tube, and whatever chemical fractionation
mechanism is responsible for the low abundances observed in 'coronal'
YSOs would be counterbalanced by the continuous flow of fresh,
undepleted material in the corona. 

\begin{acknowledgements}
  
  We wish to thank C. Ceccarelli for the useful discussions and for
  providing us with material prior to publication.  GM, SS acknowledge
  the partial support of ASI and MURST. This paper makes use of
  observations obtained with \xmm, an ESA science mission with
  instruments and contributions directly funded by ESA Member States
  and the USA (NASA). MT is financially supported by the Japan Society
  for the Promotion of Science.

\end{acknowledgements}


\begin{thebibliography}{}

\bibitem[\protect\astroncite{{Bai}}{1979}]{bai79}
{Bai} T. 1979, Sol. Phys. 62, 113

\bibitem[\protect\astroncite{{Bontemps} et~al.}{2001}]{bak+2001}
{Bontemps} S., {Andr{\' e}} P., {Kaas} A.~A. et~al. 2001, A\&A 372, 173

\bibitem[\protect\astroncite{{Boogert} et~al.}{2002}]{bhc+2002}
{Boogert} A.~C.~A., {Hogerheijde} M.~R., {Ceccarelli} C. et~al. 2002, ApJ 570,
  708

\bibitem[\protect\astroncite{{Ceccarelli} et~al.}{2002a}]{cbt+2002}
{Ceccarelli} C., {Boogert} A.~C.~A., {Tielens} A.~G.~G.~M., {Caux} E.,
  {Hogerheijde} M.~R., {Parise} B. 2002a, A\&A 395, 863

\bibitem[\protect\astroncite{{Ceccarelli} et~al.}{2002b}]{cct+2002}
{Ceccarelli} C., {Caux} E., {Tielens} A.~G.~G.~M. et~al. 2002b, A\&A 395, L29

\bibitem[\protect\astroncite{Chen et~al.}{1995}]{cml+95}
Chen H., Myers P.~C., Ladd E.~F., Wood D. O.~S. 1995, ApJ 445, 377

\bibitem[\protect\astroncite{Chiavassa et~al.}{2005}]{cct+2005}
Chiavassa A., Ceccarelli C., Tielens X., Caux E., Maret S. 2005, A\&A  in press

\bibitem[\protect\astroncite{Doppmann et~al.}{2003}]{djw2003}
Doppmann G.~W., Jaffe D.~T., White R.~J. 2003, AJ 126, 3043

\bibitem[\protect\astroncite{Favata et~al.}{2003}]{fgm+2003}
Favata F., Giardino G., Micela G., Sciortino S., Damiani F. 2003, A\&A 403, 187

\bibitem[\protect\astroncite{Favata et~al.}{2001}]{fmr+2001}
Favata F., Micela G., Reale F., Sciortino S. 2001, A\&A 375, 485

\bibitem[\protect\astroncite{Favata \& Schmitt}{1999}]{fs99}
Favata F., Schmitt J. H. M.~M. 1999, A\&A 350, 900

\bibitem[\protect\astroncite{{George} \& {Fabian}}{1991}]{gf91}
{George} I.~M., {Fabian} A.~C. 1991, MNRAS 249, 352

\bibitem[\protect\astroncite{{Imanishi} et~al.}{2001}]{ikt2001}
{Imanishi} K., {Koyama} K., {Tsuboi} Y. 2001, ApJ 557, 747

\bibitem[\protect\astroncite{{Imanishi} et~al.}{2003}]{int+2003}
{Imanishi} K., {Nakajima} H., {Tsujimoto} M., {Koyama} K., {Tsuboi} Y. 2003,
  Publ. Astr. Soc. Japan 55, 653

\bibitem[\protect\astroncite{{Imanishi} et~al.}{2002}]{itk2002}
{Imanishi} K., {Tsujimoto} M., {Koyama} K. 2002, ApJ 572, 300

\bibitem[\protect\astroncite{{Muzerolle} et~al.}{1998}]{mhc98}
{Muzerolle} J., {Hartmann} L., {Calvet} N. 1998, AJ 116, 2965

\bibitem[\protect\astroncite{{Ozawa} et~al.}{2005}]{ogm2004}
{Ozawa} H., {Grosso} N., {Montmerle} T. 2005, A\&A in press

\bibitem[\protect\astroncite{Tsuboi et~al.}{1998}]{tkm+98}
Tsuboi Y., Koyama K., Murakami H. et~al. 1998, ApJ 503, 894

\bibitem[\protect\astroncite{Tsujimoto et~al.}{2004}]{tfg+2004a}
Tsujimoto M., Feigelson E., Grosso N. et~al. 2004,
\newblock in F. Favata, G. Hussain (eds.), Cool Stars, Stellar Systems and the
  Sun, ESA-SP 560,  in press

\bibitem[\protect\astroncite{Tsujimoto et~al.}{2005}]{tfg+2004}
Tsujimoto M., Feigelson E., Grosso N. et~al. 2005, ApJS in press

\end{thebibliography}
\end{document}